\documentclass[a4paper,11pt]{article}
\pdfoutput=1 

\usepackage{jheppub} 

\usepackage[T1]{fontenc} 
\usepackage{amssymb}

\title{\boldmath Data-driven Estimation of Background Distribution through Neural Autoregressive Flows}

\author[]{Suyong Choi, Jaehoon Lim, Hayoung Oh}

\affiliation[]{Korea University, Department of Physics, Seoul 02841, Republic of Korea}

\emailAdd{suyong@korea.ac.kr}
\emailAdd{chaosbringer@korea.ac.kr}
\emailAdd{alice0102@korea.ac.kr}

\abstract{
We report on a general and automatic data-driven background distribution shape
estimation method using neural autoregressive flows (NAF), which is one of the deep generative learning methods. Data-driven background estimation is indispensable for many analyses
involving complicated final states where reliable predictions are not available.
NAF allow us to construct general bijective transformations that operate on 
multidimensional space, out of finite number of invertible one-dimensional functions. 
Given its simplicity and universality, it is well suited to the application in the data-driven
background estimation, since data-driven estimations can be expressed as transformations.
In a data-driven background estimation, the goal is to derive appropriate transformations
and apply extrapolated transformations to the region of interest. 
In the ABCDnn method, 
we can have the NAF learn the transformations' dependence on control variables  by having multiple control regions. We demonstrate that the prediction through ABCDnn method is similar to
optimal case, while having smaller statistical uncertainty.}

\begin{document} 
\maketitle
\flushbottom

\section{Introduction}
\label{sec:intro}

The LHC experiments have collected about 5\% of the total amount expected during
their lifetime. However, more data does not necessarily translate to
improved sensitivity or precision, as reducing systematic uncertainty is 
not trivial, one of which is typically due to background estimation. 
Precise estimations of the backgrounds 
are becoming increasingly important as rarer processes are being probed. 
Recent $H\rightarrow \mu^+\mu^-$
results point to the sophistication in background estimation required 
for optimal signal extraction of a very rare process,
once thought to require much more data to probe\cite{higgs2muatlas, higgs2mucms}.
With more data, more complicated final states become accessible 
and will become the focus of the experiments, since it
offers new possibilities in probing of new interactions \cite{VVV}.
It is especially for these multiparton final states where the background
estimations are difficult.

To obtain a more precise background estimation, we would need combination of higher-order
purturbative calculations from theory as well as improved understanding of
experimental apparatus and the hadron collision environment, and more advanced
analysis techniques. However, for many background processes, next-to leading order perturbative
calculations are not available, and even for those that are, the uncertainty
is significant compared to experimental uncertainties \cite{ttbbNLO,ttbb,ttbb2}.
These may limit the precision of measurements as well as some searches
for physics beyond the standard model. While having more statistics will improve some of these
limitations, we still need even higher order calculations of many background processes and improved simulations of the experimental apparatus. 

The Monte Carlo calculations are essential tools to calculate some of the essential elements of analyses. However, the calculations and simulations must confront the data in a
control regions. Differences between the simulation and the real data, in a well-controlled
phase space, are used to derive calibrations and corrections to the simulation.
Any remaining differences may be a source of experimental systematic uncertainties. 
Data-driven techniques are devised to take these differences into account when
estimating backgrounds in the region of interest.
One way to reduce the systematic uncertainties associated with 
background estimation is by making use of the large statistics data.
Historically, data-driven methods of background estimation, have been essential in bringing about the observations of many new particles \cite{jpsi1}-\cite{higgscms}. 

Despite their wide-spread use and importance, the methods of data-driven background estimation 
have not been received much scrutiny. We have shown that by extending the existing methods,
one could get improvements in predictions compared to the regular ABCD method,
by having more control regions \cite{extendedABCD}. In that study, we derived
some formulae for optimal combination of information various control regions
for extrapolation or interpolation. We cannot derive these formulae
for arbitrary case, however, by incorporating deep learning
methods, we could make the optimal use of the information in making 
background predictions.

In this letter, we introduce a general method of data-driven background shape estimation that is 
generally applicable, as it can automatically take into consideration 
the complicated correlations among the feature variables. As a case study, we will apply it to a non-trivial example of $t\bar{t}+multijets$ background estimation.

\section{Data-driven Background Estimation as Transformation}

In this section, we show that data-driven background estimation can be
formulated as a problem in finding transformation to be
applied to some base distribution.
The problem we consider is that of estimating a distribution of
$\vec{x}$, under certain condition expressed by $\vec{c}$, which we 
label as the control variable.

\begin{equation}
    p(\vec{x}| \vec{c})
\end{equation}
The control variable $\vec{c}$ can be used to label different regions of phase space,
such as various control regions (CR) and signal regions (SR). We assume
that these regions are non-overlapping. A component of $\vec{c}$ 
could be a real number, if $p$ is to depend on a 
continuous variable, or it could be an integer if it is used to enumerate different
regions of phase space.

The CRs are usually neighbors or next-to-nearest neighbors of 
the SR multidimensional space spanned by $\vec{c}$. If the CR
completely surrounds the SR then 
the problem is that of estimating $p$ through interpolation from the distributions
in the surrounding CRs.
While if signal region cannot be surrounded completely by CRs, then we need
to  extrapolate out to the SR. The so-called ``ABCD'' or ``matrix'' method frequently
employed in hadron collider experiments can be used \cite{matrixmethod} here.
Many variations on the idea are possible, from purely data-driven with no dependence
on MC, to deriving corrections to MC from the CRs.

To employ these methods,
require us to find two independent variables as control variables.
With many control regions, such requirements can be relaxed to some degree and analytic
expressions for how to make optimal extrapolations are available 
for some configurations of CR and SR \cite{extendedABCD}. 
However, for a general case, we propose to use deep neural networks
for data-driven extrapolations, as it should be able
learn the non-trivial correlations among the variables (feature or control variables)
and make the most effective use of the information available.

We formalize various data-driven techniques as transformations. 
Starting from some base
distribution $f(\vec{x}; \vec{c})$  and with some transformation $\mathcal{T}$, we can
obtain a new distribution.
\begin{equation}
    p'(\vec{x}'|\vec{c}') = \sum_{\vec{c}\neq\vec{c}'} \int \mathcal{T}(\vec{x}' ; \vec{x} | \vec{c}' ; \vec{c}) p(\vec{x}| \vec{c}) d\vec{x}.
\label{eq:gentransf}
\end{equation}
The transformation would transform from $\vec{x}$ space to $\vec{x}'$ given the condition variable $\vec{c}$, the condition under which the base distribution is obtained.
For simplicity, we will write the convolution operation and summation over $\vec{c} $ as $\mathcal{T}\otimes f$ from now on. 
All forms of data-driven techniques can be formulated in this manner.

For example, Monte Carlo simulated data distributions are compared to the real data in some
background dominated control region $\vec{c}_{b}$ and then scale factors or calibrations are derived 
\begin{equation}
    p_{data}(\vec{x}'| \vec{c}_{b}) = \mathcal{T}(\vec{x}';\vec{x}| \vec{c}_b) \otimes p_{MC}(\vec{x}| \vec{c}_{b})
\label{eq:scalefac}
\end{equation}
and applied
to Monte Carlo distribution in a desired signal region $\vec{c}_{s}$ as,
\begin{equation}
    \hat{p}_{data}(\vec{x}'| \vec{c}_s) = \mathcal{T}(\vec{x}';\vec{x}|\vec{c}_b) \otimes p_{MC}(\vec{x}| \vec{c}_s).
\end{equation}
One of the assumptions of the data-driven techniques is that $\mathcal{T}(\vec{x}';\vec{x}|\vec{c}_s) \approx \mathcal{T}(\vec{x}';\vec{x}|\vec{c}_b)$, hence $\hat{f}_{data}\approx {f}_{data}$.
 Such assumptions can be checked to some degree by using simulated data. 
If there are enough CRs then the $\mathcal{T}$ dependence on $\vec{c}$ could be learned.

\section{Neural Autoregressive Flows for Data-driven Shape Estimation}
\subsection{Neural Autoregressive Flows}
Through deep generative methods, it is possible to approximate $p(\vec{x};\vec{c})$
by training with data directly.
A powerful way for deriving transformations is the normalizing flows
method \cite{nflows1, nflows2, nflowssurvey}. Through normalizing flows, feature variables are transformed
through multidimensional invertible bijections. Such invertible bijections can be built 
using simple transformations. NF allow for probability density estimations and/or generation of variables that follow certain complicated distributions.

In this study, we adopt neural autoregressive flows, since it is simple,
but allows for universal transformation \cite{naf}.
In NAF, arbitrarily complicated transformations are created 
through a finite number of universal 1 dimensional transformations. 
This is made possible by the fact that 
multidimensional invertible bijection can be constructed in an autoregressive manner:
\begin{equation}
f(x_1,\ldots,x_d)=f_1(x_1)f_2(x_2|x_1)\ldots f_t(x_d|x_1,\ldots,x_{d-1}),
\end{equation}
where each $f_j$ is invertible in $x_j$.

In NAF, the invertible transformations are built
sequentially using a monotonic one dimensional function of $x_i$ as,
\begin{eqnarray}
y_1 & = & \hat{f}_1\left(x_1; \theta_1(\vec{c}_0)\right) \\ \nonumber
y_2 & = & \hat{f}_2\left(x_2; \theta_2(\vec{c}_0, x_1)\right) \\ \nonumber
& \ldots & \\ \nonumber
y_d & = & \hat{f}_t\left(x_t; \theta_t(\vec{c}_0, x_1, \ldots, x_{t-1})\right),
\end{eqnarray}
where $t=1,\ldots, d$. The monotonic functions $\hat{f}_t$'s are neural
networks where $\theta_t$'s are the weights and biases
of the network that depend on the previous inputs. 

An invertible one dimensional function can be built with DNN's
by using the sigmoidal function\note{$\sigma(x)= 1/(1+e^{-x})$}
and its inverse, $\sigma^{-1}$, as:
\begin{equation}
    \hat{f_i}(x_i;\theta_i) = \sigma^{-1}\left[\vec{W}^T(\theta_i)\cdot \sigma(\vec{a_i}(\theta_i) x + \vec{b_i}(\theta_i))\right],
    \label{eq:dsf}
\end{equation}
where $\vec{a}_i$, $\vec{b}_i$, $\vec{W}_i$ are $h_i$-dimensional vectors ($h_i$ is a 
hyperparameter). The monotonicity is guaranteed if all elements of $\vec{a}_i$ are positive. This is dubbed the deep sigmoidal flows (DSF) architecture.
Despite its deceptive simplicity, DSF are shown to be universal approximators
to any bijective transformations in real space \cite{naf}.

\subsection{NAF for shape estimation}

We incorporate the DSF form of NAF for a restricted form of Eq. \ref{eq:gentransf},
with the effect of summation absorbed into $p_{source}$  \cite{tf21, githubsource}.
\begin{eqnarray}
    p_{target}(\vec{x}'|\vec{c}') & =  &\int \mathcal{T}(\vec{x}' ; \vec{x} | \vec{c}' ; \vec{c})  p_{source}(\vec{x}| \vec{c}) d\vec{x} \nonumber \\
    & = & \mathcal{T}(\vec{x}' ; \vec{x} | \vec{c}' ; \vec{c}) \otimes p_{source}(\vec{x}| \vec{c})
\label{eq:naftransf}
\end{eqnarray}
In order to apply the NAF method as normalizing flows, one of the probability 
densities, either $p_{source}(\vec{x})$ or $p_{target}(\vec{x}')$ for some $\vec{c}$ must be known explicitly.
However, in our case, neither of them is known analytically.
Therefore, we use as the loss function to be minimized, the maximum-mean-discrepancy (MMD) which
is a convex function suitable for comparing finite samples
from two different multidimensional distributions \cite{MMD}. 

\begin{figure}
    \centering
    \includegraphics[width=\linewidth]{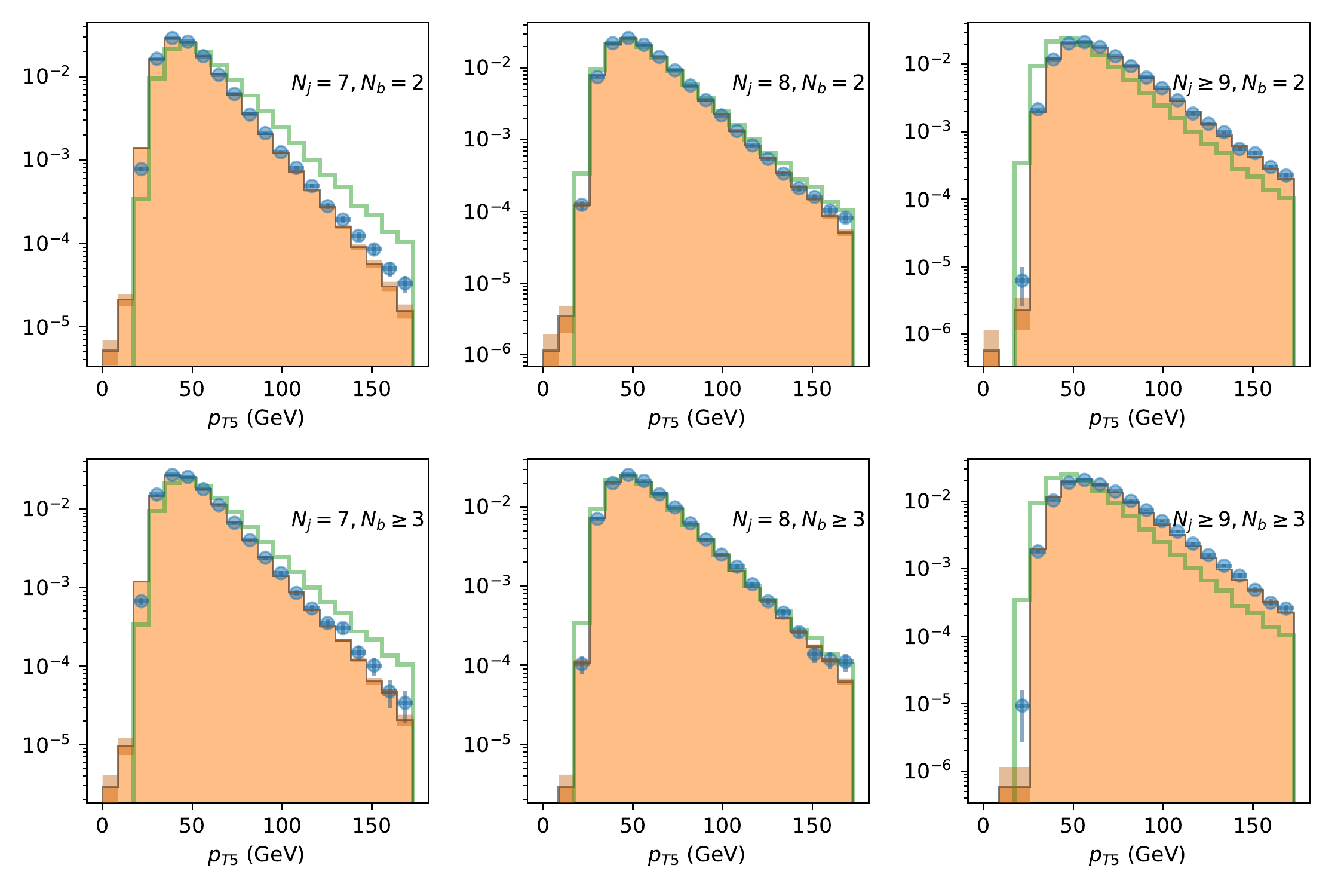}
    \caption{Distributions of the fifth leading jet $p_T$ in different regions. Open histogram is the distribution for
    all the CR's combined. The NAF morphing is applied to match the original distribution (open histogram) to the real data in each region (points), and subsequent transformed distribution (solid histogram). }
    \label{fig:pt5_log_matrix}
\end{figure}

The vectors $\vec{a}_i$, $\vec{b}_i$, $\vec{W}_i$ of Eq. \ref{eq:dsf} are
output by multilayer perceptrons. And to satisfy the requirement of $a_{ij}>0$, the activation function
at the output used is the softplus function\note{$\log(1+e^x)$}. And for $\vec{W}$,
since $W_{ij}>0$ and $\sum_j W_{ij}=1$, softmax activation is used. For $\vec{b}$,
no activation is used. The activation function for
the hidden layers is $x\sigma(x)$, also known as the ``swish'' function.
Compared to the often used RelU function, this function is continuously differentiable
everywhere and seems to provide the best performance in terms of the fidelity
of the transformed distributions.

During training, minibatches are sampled from the source and target samples, 
picked randomly from the respective subsamples in the category $\vec{c}$ and $\vec{c}'$. 
With the minibatch training, it is not possible
to consider the absolute differences in the number of events. 
Therefore, with the current method, the shape prediction is possible but not the absolute
normalization directly. However, it can be derived by adapting the method,
since it is another differential distribution. 
Also, the extended ABCD method can be used for this purpose \cite{extendedABCD}.

In the context of background estimation, the $\vec{c}$ and $\vec{c}'$ are used
to label various control or signal regions. 
The CRs neighboring SR would have more similar distributions the closer they are to the SR.
The distributions can be transformed to look like one another. 
Whether this can be done
depends on how quickly the distribution $f(\vec{x};\vec{c})$ changes depending on the condition $\vec{c}$. 
The premise of data-driven estimation is that background properties in the signal region 
can be inferred by interpolated or extrapolated from the information in the various CRs.

\begin{figure}
    \centering
    \includegraphics[width=\linewidth]{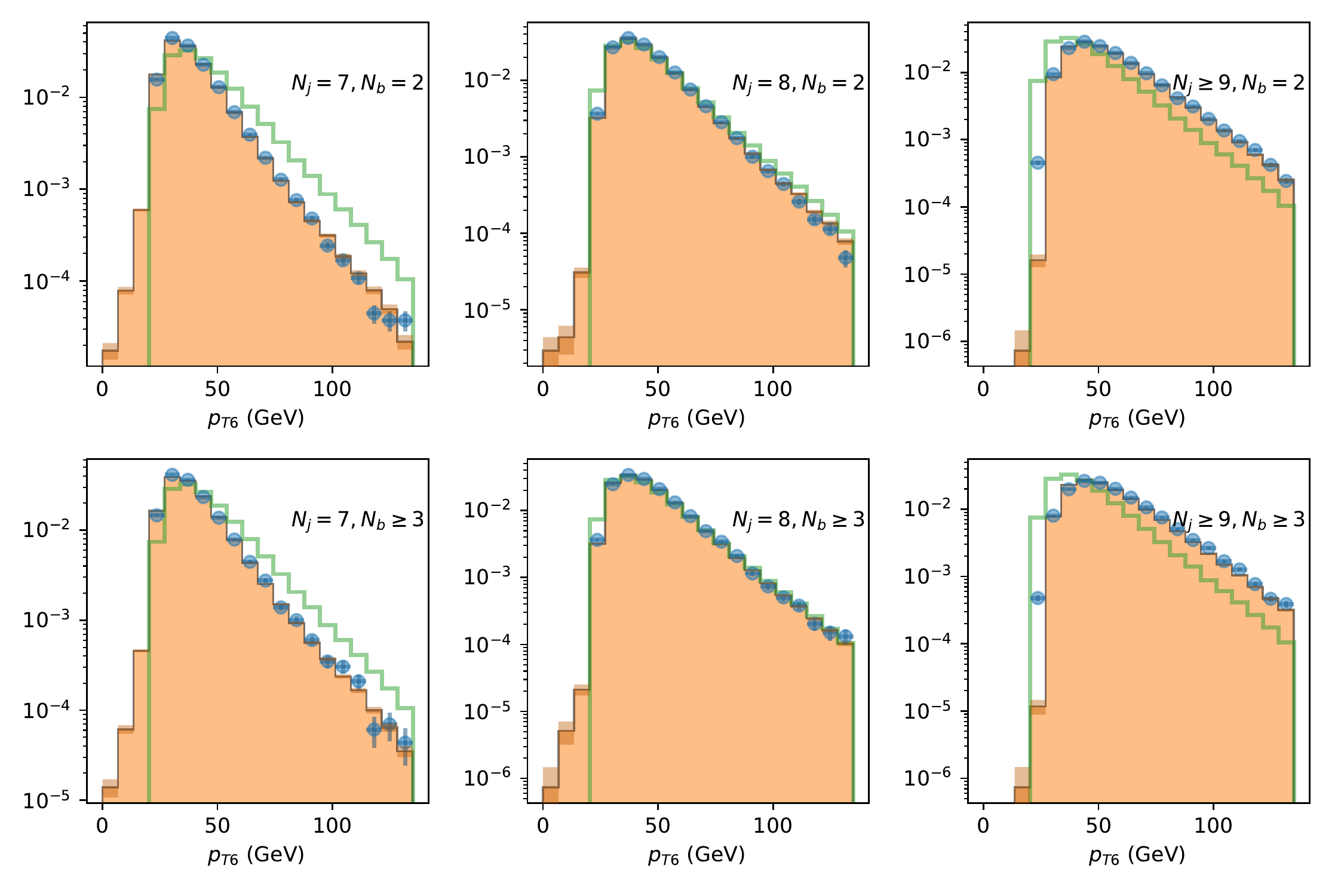}
    \caption{Distributions of the sixth leading jet $p_T$ in different regions.}
    \label{fig:pt6_log_matrix}
\end{figure}

We could think of two ways to use these transformations for background predictions. One way
is to apply transformation to a single source distribution as 
\begin{equation}
    \hat{p}_{data}(\vec{x}';\vec{c}_i) = \mathcal{T}(\vec{x}',\vec{x}; \vec{c}_i, \vec{c}_0) \otimes p_{data}(\vec{x}; \vec{c}_0),
    \label{eq:abcdnnallcr}
\end{equation}
which we implement in Tensorflow 2.1 \cite{tf21, githubsource}.
Here, the NAF transformation learns how to transform a base distribution to each control 
region. For prediction in SR, condition variable for SR is presented to the trained network. 
This could be a purely data-driven way of background estimation. 
The validity of the method and any systematic bias could be checked with simulation.
In the next section, we demonstrate this method using simulated $t\bar{t}+multijets$ sample.
Although, here $\mathcal{T}$ is learned from the data, it could be learned from MC, and applied to data.

Secondly, the NAF transformation could  be used to derive the shape corrections
to the MC simulated sample from the CRs:
\begin{equation}
    \hat{p}_{data}(\vec{x}';\vec{c}_{i}) = \mathcal{T}(\vec{x}',\vec{x}; \vec{c}_{i}, \vec{c}_{i}) \otimes p_{MC}(\vec{x}; \vec{c}_{i}).
    \label{eq:abcdnncorr}
\end{equation}
For this, the MC and data in the same CR are sampled and presented to the DNN for training. 
In this way, only the residual differences between simulation and data are learned.
While in the first case, the changes going from CR to SR are to be extrapolated based
solely on the CR without prior knowledge on anything about the SR. The physics difference
among the various CR has to be learned by the transformation.
In the second case, the important physics is already present in Monte Carlo, and by
comparing with data in respective CR, only the residual differences to account for
lack of detector understanding and some effects of higher-order contributions not present in MC
would be learned. Therefore, in the second case, the transformation would mostly appear
as small deviations away from 1. This would be a more appealing scenario for experimentalists.

The method can incorporate ABCD method, but is more general,
since it can deal with multiple CRs or multiple control variables. And, we will label this method ``ABCDnn'' since it
uses DNN for ABCD type extrapolations.
An added benefit of the ABCDnn method is that
the transformation is interpretable since the transformation on
each variable is a one dimensional function, which allows for further 
investigations.

The ABCDnn method would not be able to derive corrections for individual processes, but 
for the purpose of background estimation, it is less important.
However, if desired, it would be still possible to incorporate different transformations to different
samples by small modifications to the methods. 
For example, one could create a sample of MC with appropriate admixture of well
understood backgrounds and another sample with larger uncertainties. During the
training, we can choose minibatch and apply the transformation only to the less
understood sample.

\section{Application of ABCDnn to $t\bar{t}+multijets$}
In this section, we apply Eq. \ref{eq:abcdnnallcr} to $t\bar{t} + multijets$ simulated data. In our previous
study, we found an analytic expression for an optimized method of extrapolation
extending the ABCD method for the case of multiple CRs under some assumptions \cite{extendedABCD}.
In the ABCD method, the two variables should be independent. With multiple CRs, 
non-linear dependence of the distributions on control variables
and some effects of correlations among control variables can be reduced.

Through this case study, we would like to understand firstly, whether ABCDnn method can be used
for extrapolations to SR estimate the distributions of various kinematic variables, and secondly, 
how it compares with the analytically derived extended ABCD method and whether
it offers any advantages.

\begin{figure}
    \centering
    \includegraphics[width=\linewidth]{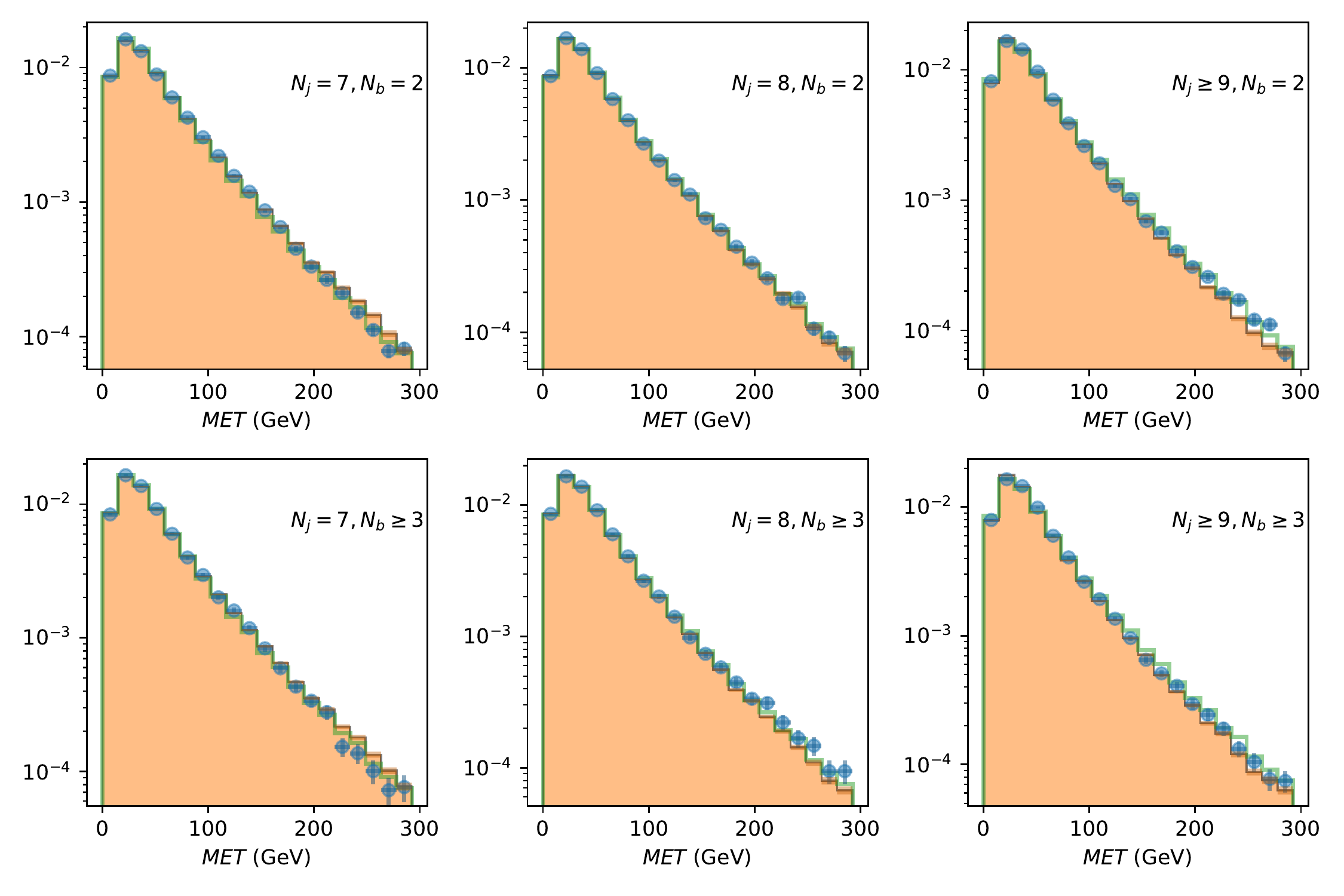}
    \caption{Distributions of missing $E_T$ in different regions.}
    \label{fig:met_log_matrix}
\end{figure}

\begin{figure}
    \centering
    \includegraphics[width=\linewidth]{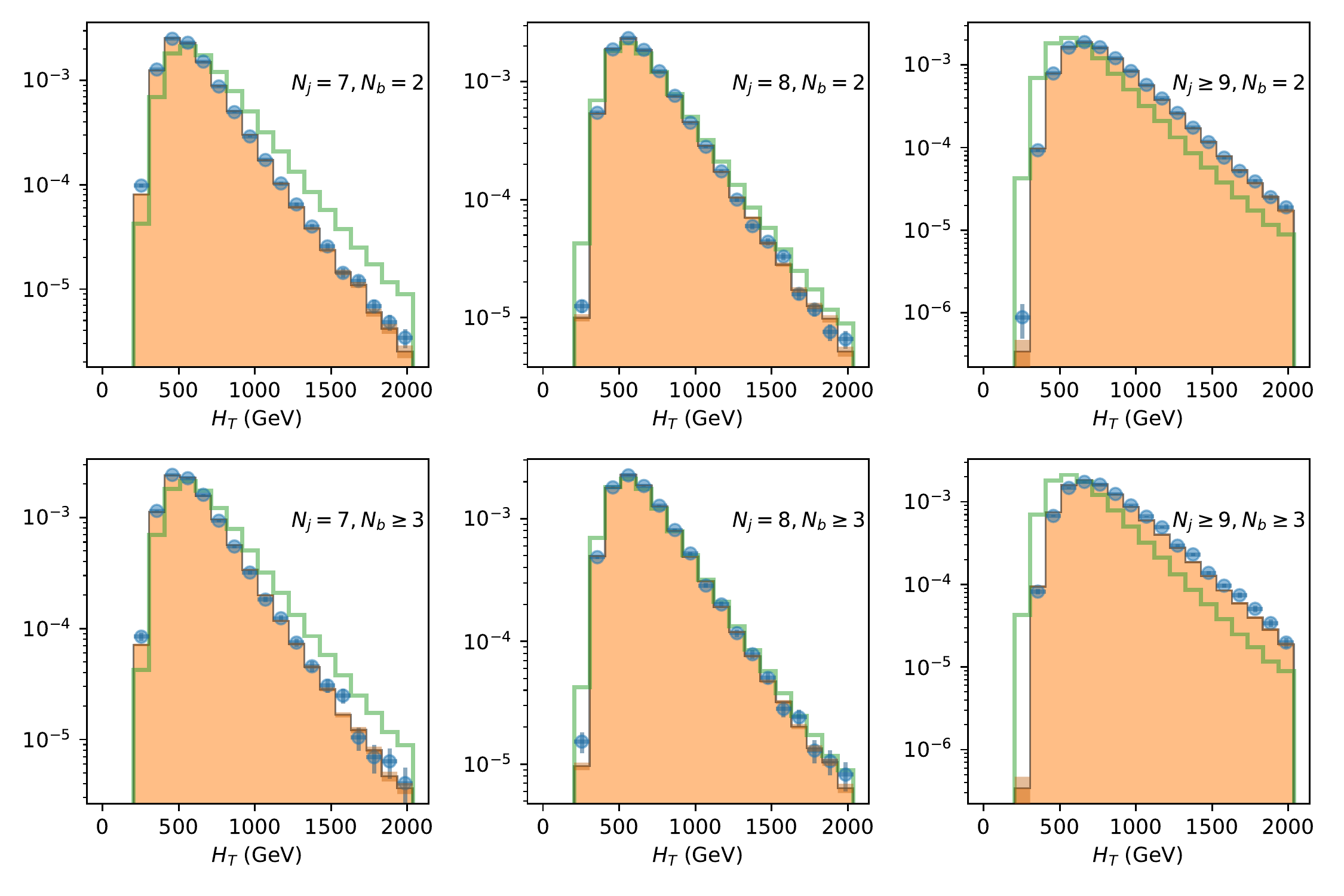}
    \caption{Distributions of $H_T$ in different regions.}
    \label{fig:ht_log_matrix}
\end{figure}

Data sample in this study was generated with MadGraph5 \cite{mg5}. 
Process generated was $t\bar{t}+jj$ at leading order, where the $W$ bosons from the top decays were forced to decay hadronically. This process was chosen as a proxy for backgrounds to many searches 
involving many jets in the final states. 
The generated sample was subsequently parton-showered and hadronized
with Pythia 8 \cite{pythia8}. Finally, fast detector simulation and object reconstruction
was done using Delphes 3 with the default settings \cite{delphes3}.
The hadronic jet cone size used was $\Delta R=0.4$, and the b-jets were
identified through parameterized b-tagging efficiency and fake rates 
implemented in Delphes 3.

The control variables chosen were the number of hadronic jets ($N_{j}$)
and the number of b-tagged jets in an event ($N_{b}$). For the baseline selection,
we required $N_{j}\geq 7$ and $N_{b}\geq 2$. The signal region (SR) chosen was $N_{j}\geq 9$ and $N_{b}\geq 3$, and the remaining regions are CR's. We used the form \ref{eq:abcdnnallcr} where 
we used the data sample in all the CR's together to form the source distribution.

The training was done by using minibatch scheme. We need two minibatches per training step,
one from the target distribution and the other from the source distribution corresponding to $\vec{c}_0$.
The source minibatch is made by randomly sampling from data 
from all the CRs combined.
And for the target minibatch, we first select the specific CR i.e. ($N_j$ and $N_b$) 
 by random sampling from all the CRs combined. Then, we randomly select events from that 
specific CR, so that samples within a minibatch have the same value of $\vec{c}$.  
Due to the discrete nature of the condition variables, they
were ``one-hot'' encoded such that deep neural networks do not have difficulty 
in making use of this information. For more details on the setup, consult \cite{githubsource}.

Figures \ref{fig:pt5_log_matrix} - \ref{fig:ht_log_matrix} show
some of the kinematic variables in $t\bar{t}+mulijets$ sample. 
In each plot, the open histogram is the distribution of all CRs combined (the source),
and they are identical across the six plots in each figure.
The learned transformation is applied to the source data in CR which
are shown as solid histograms. And the actual distributions (the target) are shown as solid points. 
The rightmost bottom plot is the SR. To make the prediction in SR,
the condition variable for the SR is presented to the NAF transformation together
with the source data for this purpose. 

We can see that the transformations for each CR is properly learned.
Also, for the SR, the shapes are well reproduced by the transformation. 
We emphasize that this transformation is not done on a variable by
variable basis, but to all the feature variables simultaneously for a given event. 
Unless the correlations among variables are considered correctly it is 
extremely unlikely to obtain such agreement.
We note that the transformed distribution has less fidelity in the region where
there is a sharp cut off, such as near the $p_T$ threshold of the hadronic jets.
The hadronic jets were selected to have $p_T>20\mathrm{GeV}$, but the transformed $p_T$
sometimes crosses this boundary, albeit at a very low rate. 
In practice, such events will be thrown away or assigned some systematic uncertainties
for shape prediction.

By comparing the six plots in each figure, we can glimpse that the actual distributions,
marked with solid points, show systematic trend as $N_j$ or $N_b$ changes, and 
this trend is what is learned by the NAF transformation when $N_j$ and $N_b$
is used as control variables. Taking Fig. \ref{fig:pt5_log_matrix} as an example,
compared to the open histogram, the actual distributions become progessively
harder as $N_j$ increases. The spectra also become harder as $N_b$ increases, by
comparing the solid points in the top panes to those in the bottom panes, but to much
lesser degree than $N_j$. If there is a component of the background that shows
completely novel property in SR and which cannot be extrapolated from the CRs
then this method would not be able to predict such novel feature. But this is expected.

\begin{figure}
    \centering
    \includegraphics[width=0.45\linewidth]{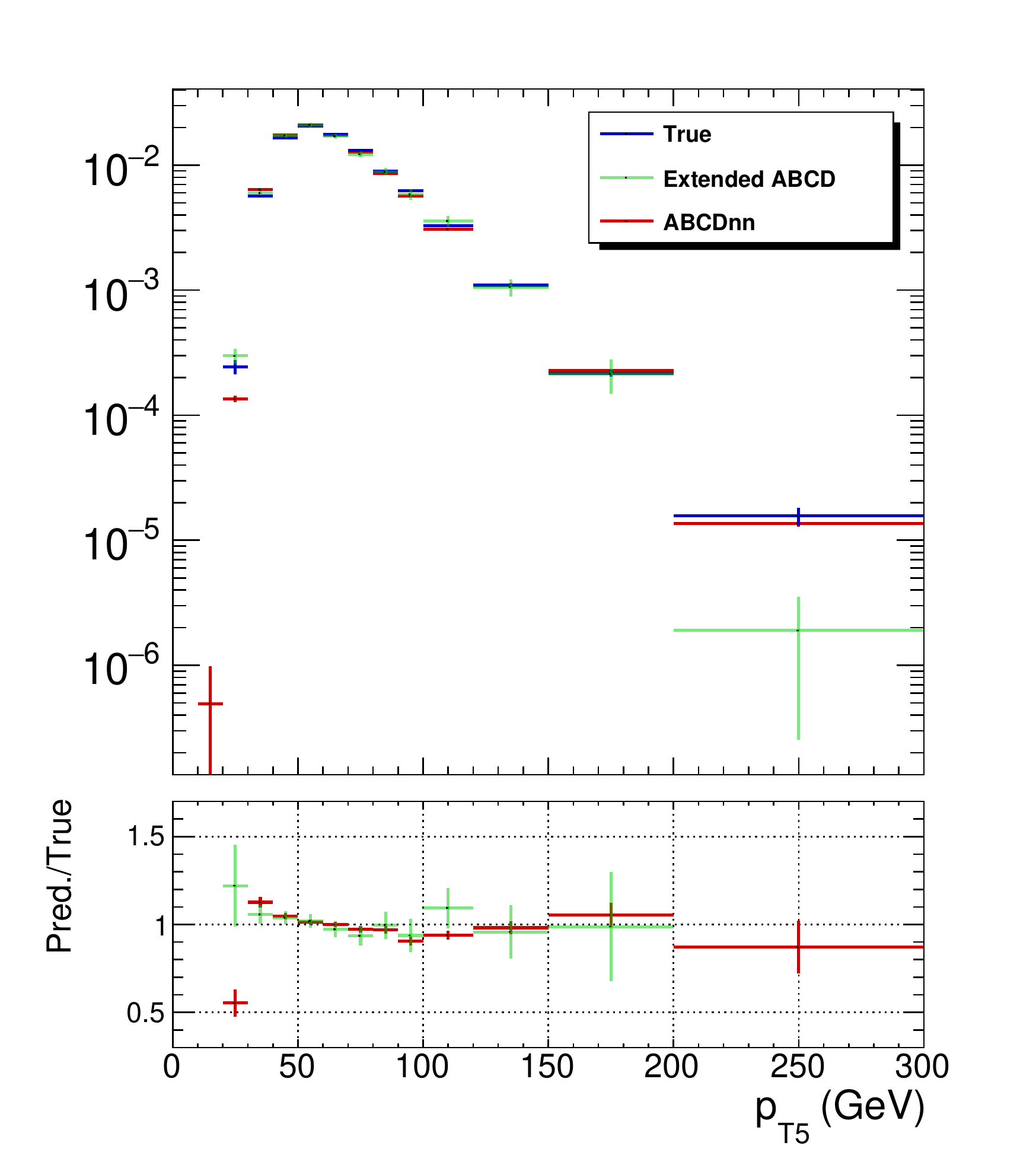}
    \includegraphics[width=0.45\linewidth]{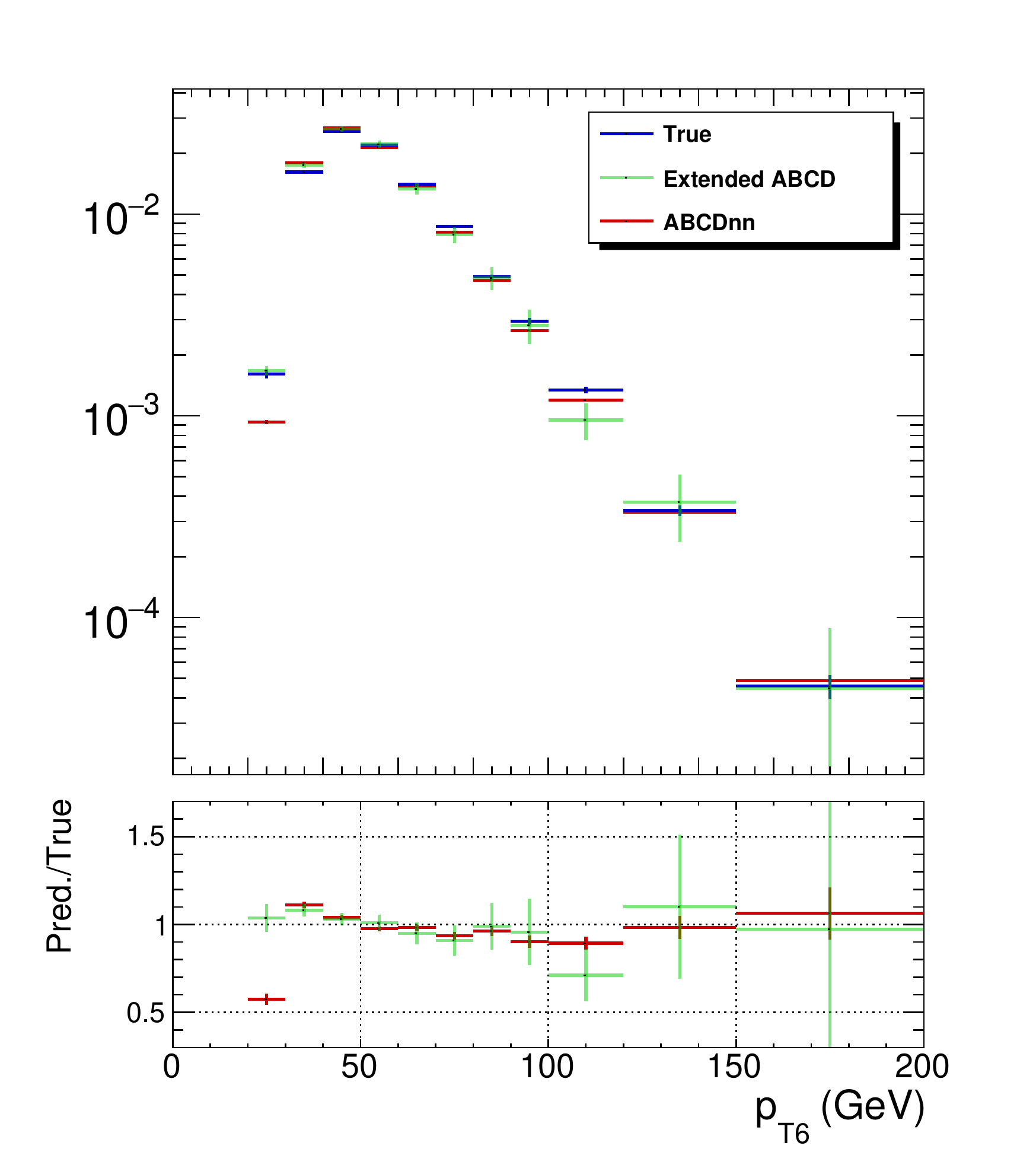}
    \caption{Normalized $p_T$ distribution of the fifth and sixth leading jet in $N_{jet}\geq 9$ and $N_{bjet}\geq 3$ and predictions through the extended ABCD (green error bars) and ABCDnn (red error bars) methods}
    \label{fig:pt59}
\end{figure}

\begin{figure}
    \centering
    \includegraphics[width=0.45\linewidth]{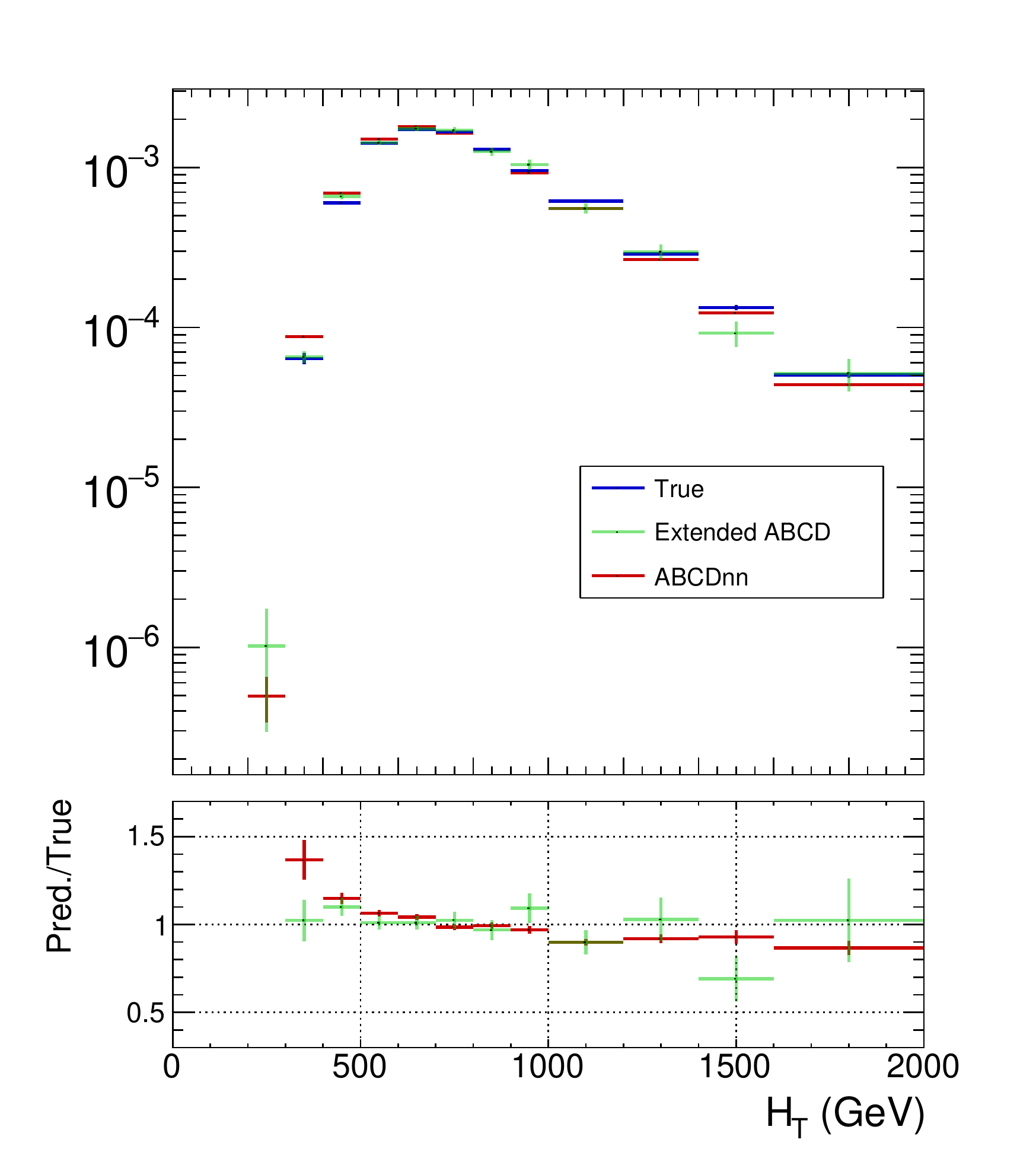}
    \includegraphics[width=0.45\linewidth]{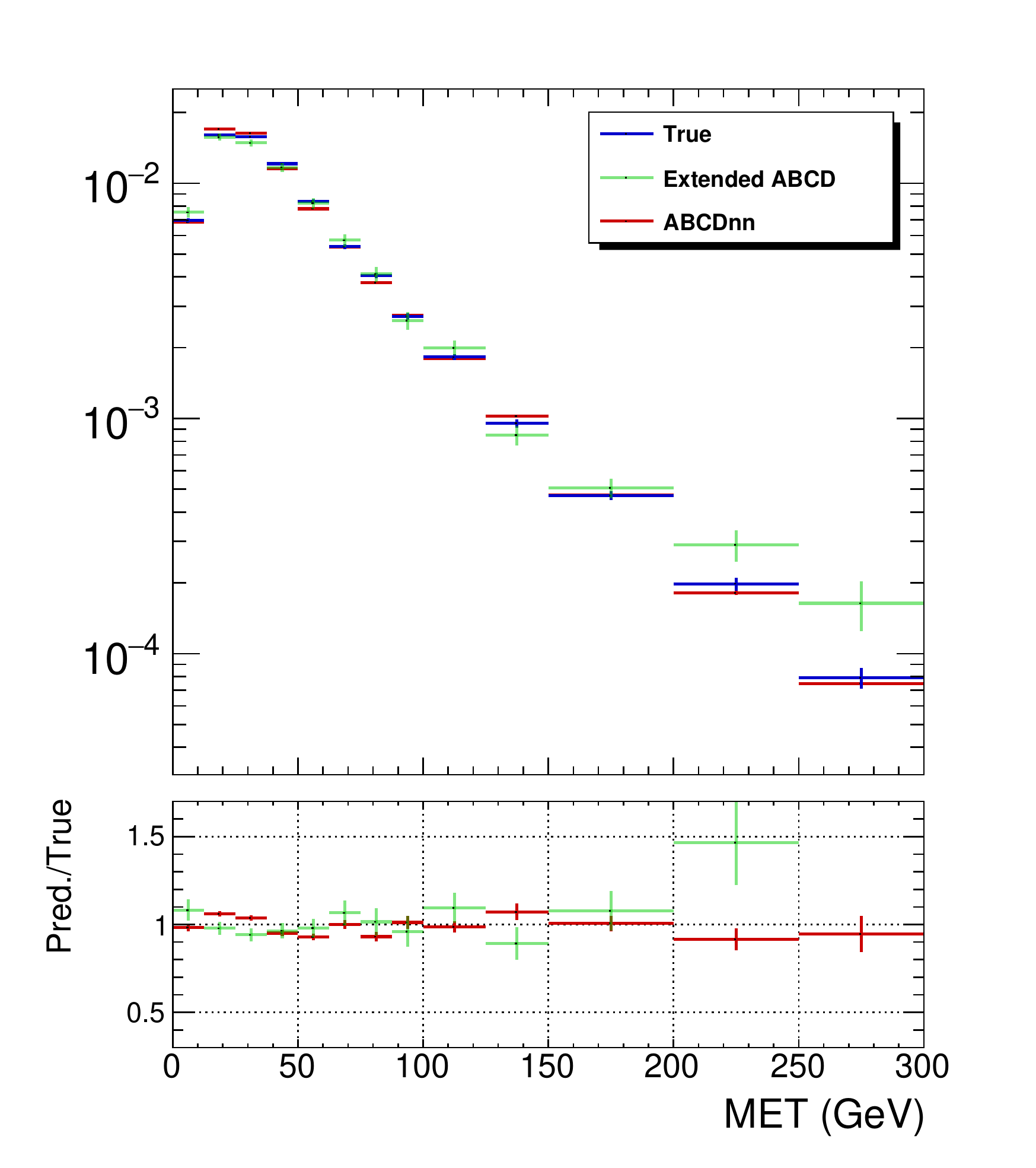}
    \caption{Normalized distribution of $H_T$ and missing $E_T$ of $t\bar{t}+multijets$ in SR ($N_{jet}\geq 9$ and $N_{bjet}\geq 3$) and predictions through the extended ABCD (green error bars) and ABCDnn (red error bars) methods.}
    \label{fig:ht9}
\end{figure}

In Figs. \ref{fig:pt59}-\ref{fig:ht9}, we show distributions in the SR only and 
ratios to the true distributions, which illustrates the quality of the predictions
with ABCDnn. We also compare them with the predictions made using the extended ABCD method \cite{extendedABCD}. 
The extended ABCD method makes use of distributions in multiple CR's and makes predictions in the SR by taking optimal (under some assumptions) products and divisions among the 1-D distributions in various CRs. 
By considering multilple CR's the extended ABCD
method is able to take into consideration 
some non-linear dependence of the distributions on the control variables and weak correlation
between the control variables, to some degree. The general tendencies of both
predictions are similar and within statistical errors for the most part.

In each method, all the available statistics in CR is used for prediction in SR.
However, the extended ABCD method can only be applied to one variable at a time and
if some bin in one of the histogram has large statistical uncertainty then it will impact directly
the final result, since products and divisions among the histograms in various CRs are used. 
While, With the ABCDnn, multidimensional distribution is morphed in a non-linear manner by shifting
values of the feature variables, and at the same time preserving correlation. 
We observe that the extended ABCD predictions have on average larger statistical uncertainties.
Through this case study we demonstrated that the NAF transformation is able to 
learn transformations from CR and extrapolate it to SR. And interpreting the
tendency compared to the extended ABCD method, we might interpret that the behavior is
striving for optimal use of information available. Especially, ABCDnn makes the best use of the statistics available, while for extended ABCD method, the statistical uncertainty could not
be taken into account when aiming for optimal use of data.

The ABCDnn method would allow systematic and automatic approach to background estimations, 
and it can be adapted to a more variety of use cases than what is considered here.
This study was restricted to using the simulated data, in place of 
the real data, for the purpose of purely data-driven background shape estimation scenario. 
We can apply this method to a case where more than two control variables are used
without any modifications, but this requires a dedicated study.
And by modifying the method slightly, it is possible to implement  Eq. \ref{eq:abcdnncorr}, 
where only the corrections between the real data and the simulated data are learned,
which would be appealing for experimentalists in deriving corrections to
simulation and understanding their meanng. 
Although the background shape estimations was the primary
focus of this study, with further modifications, it would be possible 
to estimate the relative event rates in each categories.
(We note that while preparing the manuscript,
an idea for using DNN for ABCD extrapolation appeared \cite{ABCDisCo}.)

\section{Conclusions}
We presented a novel general data-driven method using neural autoregressive flows (NAF)
for obtaining background distributions, which we dubbed the ABCDnn method.
Through the use of multiple control regions, the ABCDnn method 
learns the dependence of transformation
on the control variables, and it is able to extrapolate/interpolate to region 
of interest that neighbors the control regions.
Since the transformation is constructed out of a finite number of 1-D transformations,
it is possible to understand or interpret it, unlike some DNN methods.
Hence, the method 
Whereas existing data-driven estimation methods usually work on a single 
feature variable and it would is able to handle simultaneously many variables automatically
while taking into account the correlations among feature variables, unlike existing
methods. Moreover, the evidence from case study suggests that the prediction is close
to optimal. This method can be used in many cases where the 
reliable predictions of backgrounds based solely on simulations is not available. 

\acknowledgments

This work has been suppored by Korean National Science Foundation
through its mid-career grant.


\end{document}